# Prospects for Superconductivity in LiBSi

O. P. Isikaku-Ironkwe[1, 2]
[1]The Center for Superconductivity Technologies (TCST)
Department of Physics,
Michael Okpara University of Agriculture, Umudike (MOUAU),
Umuahia, Abia State, Nigeria
and
[2]RTS Technologies, San Diego, CA 92122

## Abstract

The search for superconductivity in materials iso-structural and iso-electronic with magnesium diboride, $MgB_2$, has not yielded results with Tc near 39K. In particular, Lithium borocarbide, LiBC, and its modification $Li_{0.5}BC$, proved to be non-superconducting. Using our materials-specific condition for superconductivity and the symmetry principles for superconductor search, we show that LiBC cannot be a superconductor. By replacing C in LiBC with Si we achieve valence electron count (Ne) and atomic number (Z) symmetry with $MgB_2$. We therefore predict that Lithium boron silicon, LiBSi, should be a superconductor with Tc comparable to $MgB_2$.

## Introduction

The search for superconductivity in materials iso-structural and iso-electronic with $MgB_2$ [1] has been intensively and extensively pursued theoretically [2 - 5] and experimentally [6 - 14] without yielding results with Tc near 39K. In particular, LlBC predicted [3] to be a better superconductor than $MgB_2$ proved not to be superconducting at all [6 - 9]. Investigating how to improve these results [15 - 19], led to the elucidation of material specific parameters that determine transition temperature Tc, namely electronegativity, valence electrons, atomic number and formula weight. In this paper, we apply the Tc formula and symmetry rules earlier proposed [18, 19] to explain why LiBC is not superconducting and propose an alternative material, LiBSi, which we show to be superconducting by the symmetry rules. We also compute its Tc.

## Basic Concepts

We showed in [18] that the transition temperature, $T_c$, of a superconductor can be estimated in material specific properties of electronegativity, $x$, valence electrons, Ne, atomic number, Z, and a parameter, Ko by the equation:

$$T_c = x \frac{Ne}{\sqrt{Z}} K_o \qquad (1)$$

where $K_o$ is a parameter that determines the value of Tc. Ko = n(Fw/Z) and n is dependent on the family of superconductors. Fw represents formula weight of the superconductor. For $MgB_2$, with Tc of 39K and Fw/Z of 6.26, Ko = 22.85, making n = 3.65. Computation of average values of $x$, Ne and Z are also shown in [18] and displayed in Table 1 for the relevant materials.

## Symmetry Principles
In our earlier papers [18, 19], we stated the symmetry rules that determine superconductivity in similar materials. Applying these symmetry rules to LiBC and $MgB_2$, we find that LiBC does not match $MgB_2$ and also that the Ne/Sqrt Z for LiBC is greater than 1, which forbids superconductivity. From the Periodic Table, Si and C are in the same group, with silicon heavier than carbon. What if we replace C with Si? Will LiBSi form? Will it meet the symmetry rules? For LiBSi, we find that it has the same valence electron count, Ne, and same atomic number, Z, as magnesium diboride (Table 1.) The electronegativity of LiBSi is 1.6 and $MgB_2$'s is 1.7333. Also, as shown in table 1, their formula weights are almost the same. The second symmetry rule [18, 19] states that their Tcs will be proportional to their electronegativities. Since 0.8< Ne/$\sqrt{Z}$ <1.0, we expect superconductivity in LiBSi with Tc of (1.6/1.7333) of 39K, that is 36K, assuming 2-gaps like $MgB_2$. Otherwise we get a Tc of 19.5K

## Discussion
In the search for magnesium diboride like superconductors, we have come to learn that structural similarity and iso-valency are not sufficient conditions [19]. The material must pass the superconductivity test and the symmetry rules test [18, 19]. One of the first steps to follow for 3-atom $MgB_2$-like compounds is that the sum of the atomic numbers must be 22 and the sum of the electronegativities close to 5.2. This strategy has led us to very many

MgB$_2$-like materials. In the example of LiBC we see that the sum of the atomic numbers is 14. Replacing C with Si gives a sum of 22, the same as in MgB$_2$. The valence electron count remains 2.667 while the sum of the electronegativities is now 4.8, close to 5.3 of MgB$_2$. The new material LiBSi now meets a symmetry rule for superconductivity when compared to MgB$_2$. However, the unanswered question is if LiBSi can actually form. Ref. [20] suggests that LiBSi is a glass and may be prepared by ball milling for 24 hrs in ethanol, the appropriate weights of the compounds Li$_2$CO$_3$, SiO$_2$ and H$_3$BO$_3$, which are then heated to 750 degrees centigrade in Pt crucibles. LiBSi has been used as an additive to enhance the dielectric properties of certain ceramic materials. It should be noted that Li, B and S tend to form lithium borosilicate with oxygen.

## Conclusion

We have solved the problem of no superconductivity in LiBC. Just replace the carbon in LiBC with iso-valent and heavier silicon. We get LiBSi which has the same Ne and Z as MgB$_2$ and an electronegativity of 1.6. Applying the symmetry rules and the Tc equation (1), we computed the Tc of LiBSi as 36K, assuming two gaps like MgB$_2$. The prediction of superconductivity in Lithium boron silicon, LiBSi, awaits experimental verification.

## Acknowledgements

The author acknowledges enlightening discussions on symmetry in physics, chemistry, mathematics and nature with A.O.E. Animalu at the University of Nigeria, Nsukka; with M.B. Maple at UC San Diego and with J.R. O'Brien at Quantum Design San Diego. This research was supported by M.J. Schaffer then at General Atomics.

## References

1. J. Nagamatsu, N. Nakagawa, T. Muranaka, Y. Zenitani and J. Akimitsu, "Superconductivity at 39K in Magnesium Diboride," Nature 410, 63 (2001)
2. N.I. Medvedeva, A.L. Ivannovskii, J.E. Medvedeva and A.J. Freeman, "Electronic structure of superconducting MgB$_2$ and related binary and ternary borides", Phys. Rev. B 64, 020502 (R) 2001 and references therein
3. H. Rosner, A. Kitaigorodsky and W.E. Pickett, "Predictions of High Tc Superconductivity in Hole-doped LiBC", Phys Rev. Lett. 88, 127001 (2002).

4. C. Bersier, A. Floris, A. Sanna, G. Profeta, A. Continenza, E.K.U. Gross and "Electronic, dynamical and superconducting properties of CaBeSi", ArXiv: 0803.1044 (2008).
5. Hyoung Jeon Choi, Steven G. Louie and Marvin L. Cohen, "Prediction of superconducting properties of $CaB_2$ using anisotropic Eliashberg Theory", Phys. Rev. B 80, 064503 (2009) and References 1 - 21 in that paper.
6. A. Bharathi, S. Jemima Balaselvi, M. Premila, T.N. Sairam, G.L.N. Reddy, C.S. Sundar, Y. Hariharan "Synthesis and search for superconductivity in LiBC" Arxiv:cond-mat/0207448V1 and references therein., Solid State Comm, (2002), 124, 423
7. Renker, H. Schober, P. Adelmann, P. Schweiss, K.-P. Bohnen, R. Heid ,"LiBC - A prevented superconductor",Cond-mat/0302036
8. A.M. Fogg, J.B. Calridge, G.R. Darling and M.J. Rossiensky "Synthesis and characterization of $Li_xBC$---hole doping does not induce superconductivity". Cond-mat/0304662v1
9. A. Lazicki, C.S. Yoo, H. Cynn, W.J. Evans, W.E. Pickett, J. Olamit, Kai Liu and Y. Ohishi "Search for superconductivity in LiBC at high pressure: Diamond anvil cell experiments and first-principles calculations" Phys. Rev. B 75, 054507 (2007)
10. I. Felner "Absence of superconductivity in $BeB_2$", Physica C 353 (2001) 11 – 13.; D.P. Young, P.W. Adams, J.Y. Chan and F.R. Franczek, "Structure and superconducting properties of $BeB_2$" Cond-mat/0104063
11. B. Lorenz, J. Lenzi, J. Cmaidalka, R.L. Meng, Y.Y. Sun, Y.Y. Xue and C.W. Chu, "Superconductivity in the C32 intermetallic compounds $AAl_{2-x}Si_x$, with A=Ca and Sr; and 0.6<$x$<1.2" Physica C, 383, 191 (2002)
12. R.L. Meng, B. Lorenz, Y.S. Wang, J. Cmaidalka, Y.Y. Xue, J.K. Meen. C.W. Chu"Study of binary and pseudo-binary intermetallic compounds with $AlB_2$ structure"Physics C: 382 113–116(2002).
13. R.L. Meng, B. Lorenz, J. Cmaidalka, Y.S. Wang, Y.Y. Sun, J. Lenzi, J.K. Meen, Y.Y. Xue and C.W. Chu, "Study of intermetallic compounds isostructural to $MgB_2$, IEEE Trans. Applied Superconductivity, Vol. 13, 3042- 3046 (2002).
14. Cristina Buzea, Tsutomu Yamashita, "Review of superconducting properties of $MgB_2$", Superconductor Science & Technology, Vol. 14, No. 11 (2001) R115-R146
15. O. Paul Isikaku-Ironkwe, "Electronegativity Spectrum Maps: A computational combinatorial materials synthesis and search tool". http://meetings.aps.org/link/BAPS.2008.MAR.K1.50
16. O. Paul Isikaku-Ironkwe, "Search for Magnesium Diboride-like Binary Superconductors" http://meetings.aps.org/link/BAPS.2008.MAR.K1.7
17. O. Paul Isikaku-Ironkwe, "Do 'magic' electronegativities exist for superconductivity?",http://meetings.aps.org/link/BAPS.2008.MAR.K1.47
18. O. P. Isikaku-Ironkwe, "Transition Temperatures of Superconductors estimated from Periodic Table Properties", Arxiv: 1204.0233 (2012)

19. O. P. Isikaku-Ironkwe, "Possible High-Tc Superconductivity in LiMgN: A $MgB_2$-like Material", Arxiv: 1204.5389 (2012)
20. Yonggang Zang, Shunhua Wu, Shungi Gao, Xuesong Wei, "Improvement in the dielectric properties of $Ba_{6-3x}(Nd_{0.4}Bi_{0.6})_{8+2x}Ti_{18}O_{54}$ (x=1-1.5) ceramics by LiBSi/BaLiF glass additive", J. Mater Sci: Materials in Electronics (2011) 22: 561 - 564

| | Material | $x$ | Ne | Z | Ne/$\sqrt{Z}$ | Fw | Fw/Z | Tc(K) | Ko |
|---|---|---|---|---|---|---|---|---|---|
| 1 | $MgB_2$ | 1.7333 | 2.667 | 7.3333 | 0.9847 | 45.93 | 6.263 | 39 | 22.85 |
| 2 | LiBC | 1.8333 | 2.667 | 4.6667 | 1.2344 | 29.6 | 6.343 | 0 | 0 |
| 3 | LiBSi | 1.6 | 2.667 | 7.3333 | 0.9847 | 45.84 | 6.251 | 35.95 | 22.85 |

Table 1: Material specific characterization datasets (MSCDs) for $MgB_2$, LiBC and LiBSi. Note that Ne and Z are the same for $MgB_2$ and LiBSi. Their Formula weights (Fw) differ by just .09. The Tc of LiBSi was computed with equation (1). The other parameters were computed as shown in Ref. [18].